# Performance Enhancement Factors of ERP Projects in a Telecom Public Sector Organization of Pakistan: An Exploratory Study

[1]Shafqat Ali Shad (shafqat@mail.ustc.edu.cn),

[2]Enhong Chen (cheneh@ustc.edu.cn)
Department of Computer Science and Technology
University of Science and Technology of China
Huangshan Road, Hefei, 230027 Anhui, China

[3]Faisal Malik Faisal Azeem (faisal.azeam@comsats.edu.pk)
COMSATS institute of information technology, Pakistan
CIIT, Quaid Avenue Wah Cantt, Pakistan

# Performance Enhancement Factors of ERP Projects in a Telecom Public Sector Organization of Pakistan: An Exploratory Study

**Abstract**

Public sector organizations are treated in a different manner, as Information technology/information system has become necessity in a highly competitive environment. Importance of information systems is becoming more and more vital as the global technology adoption is in progress. ERP projects are considered as one of the most important and critical area of technology especially in public sector organizations where cost effectiveness and operational efficiency is prioritized on profits. A lot of research studies have been made on ERP projects but mostly on critical success factors (CSFs) and other managerial issues. This ongoing research mainly focuses upon the performance of ERP software in public sector organizations by thoroughly going through one of the largest most public sector organizations of Pakistan. Though ERP projects are handled by experienced consultants in addition to the support of vendor companies but because of its features and functional complexity and mismatch with the organizational processes, different managerial and technical issues arise during and after its implementation. This study investigates the performance of ERP system in a large public sector organization of Pakistan; five most critical technical factors, which can lead the whole project towards success or failure, have been dug out. This exploratory study i.e. extensive literature review and a case study includes survey and interviews which later on investigated through their application in a public sector organization in Pakistan. Statistical analysis later on proves that the developed hypothesis which ultimately paves the way for the findings i.e. five critical technical factors, if properly addressed in addition to the other managerial factors can enhance the performance of ERP projects. Study also finds the gap between the existing literature and its real application, which derives the researchers, more towards exploration, and application of crucial technical factors.

# INTRODUCTION

ERP systems implementation is the critical most source of technological change in any large-scale organizations. Technological change when becomes inevitable, public sector organizations also need to be in line with other private sector organizations to serve their core objectives. Though ERP implementation leads the organizations towards success but the study explored facts about ERP project implementation failure. (Davenport T. H., 1998).

Existing successful ERP implementations though can be exemplified for other ongoing ERP projects but it is quite difficult to learn for software process improvement especially in a public sector organization where the organizational processes aim different. This research study will determine the most critical technical factors, which facilitate public sector organizations in performance enhancement. This paper finds the crucial technical factors determining the performance of ERP projects and provides recommendations to carry these five factors as performance indicator in addition.

Organizations where ERP implementation projects undergo require primarily process and standardization across each part of it. Package software(s) are the core facilitator of effective and efficient decision making by editing meaning to the transactional data into meaningful form i.e. analytical reports etc for not only tactical but strategic planning and decision making. Companies when ever faced technology/information system complexity issues that caused higher costs, time and reduction is productivity, thought of a new compact Information System solution(s). In ab-initio of 90's when company found ERP as a solution of their whole package of problems, they started redesigning their processes to be ready for ERP adoption. 'Business first then technology' slogan whenever neglected caused organizational collapsed. As a result in this specific century a great number of ERP implementations reported as failure. So business processes were redesigned for successful ERP systems implementations. Performing Value addition process design activities by the employees directed the organization(s) towards productivity enhancement. As the mismatch between ERP systems and organizational process diagnosed, dilemma to chose software features change or to modify business process to align these with standardized ERP system features, Organizations found it more productive, comparatively easy and less costly the later one (Davenport T. H., 1998) (Holland & Light,1999) .

Earlier research studies pointed out critical complexities and high failure rates in ERP systems implementation (Davenport T. H., 1998). One of the major reasons of this failure is shortage or of technical critical success factors studies, which could provide standardized rout towards successful ERP implementation. ERP systems have not yet been critically theorized from its important technical complexities perceptively in a sufficient manner by the researchers for ongoing ERP systems implementations (Brown & Vessey,1999) . ERP systems, as a source of successful information management tool (Fiona et al, 2001) which helped organizations to drastically change the way they utilize their resources with a novel and interlinked applications solution i.e. ERP across each sector of the organization. ERP systems facitilitate organizations in understanding the process view of their businesses to develop standardized processes (Fiona et al, 2001) . The study explores that the suggestions are given for reexamining all the processes of the business when ERP intended to be implemented (Hammer et al, 2001) . ERP system implementation projects have always been considered as the most critical and risky ones for businesses that intend to adopt it because of its width of complexities, especially the technical ones that ultimately effect on the business performance (Parr et al, 2000).

The study formulated five hypotheses in order to explore whether these complex and critical factors are on the right track towards business performance excellence. In order to determine the performance of ERP projects these factor to what extent are considered the most critical ones. ERP systems implementations though have been implemented successfully in many MNEs but when explored, found challenging because it differs from a simple set of software or information system deployments (Sumner, 2000) because of its varying set of complexities and risk factors. ERP Total cost of ownership is much more than the other developed information systems especially the time estimation for ERP projects is comparatively critical and longer (Wenhong et al, 2004). An independent survey reflects that global CIO's prioritize ERP system performance in top five IT governance issues of their organizations (Kenneth et al, 2004).

Studies, as stated earlier are present on critical success factors that address managerial issues but the critical technical issues have not yet been uncovered in a required extend through the real life exploration. Though some tried to establish technical critical factors which impact on ERP systems performance but this study goes in deep of most critical ones which do have impacts on ERP performance not only from literature but through a real term exploration of a public sector organization issues. Case study is a well-known method of research for exploratory theory building research (Yin RK, 1989). In addition to the review of the literature this exploratory case study consisted of a survey from a public sector organization of Pakistan, which is being transformed from public to private mainly by having BPR & ERP system (SAP) implementation. Though the company is in the process of restructuring but the study limits itself on five number-technical critical factors of ERP system performance.

The ultimate results because of being concluded from the two different methodologies followed in this paper i.e. from literature review and exploratory case study which would provide empirical data analysis from the survey in one of the largest telecom concern where ERP system has successfully been implemented.

The first methodology includes a case study which has been derived from the survey and the interviews from 58 employees in a large scale telecom public sector organization which is being guarded in technology especially by ERP i.e. SAP system implementation.

Results from this survey and interview have been derived by both qualitative and quantitative data analysis that leads the researcher towards the ultimate findings. Four statistical hypotheses were formulated and statistical tests applied provided a favorable support towards these hypotheses. The results provide the study that these crucial factors do have a great impact on ERP projects performance if catered for rightly and on the right time. In addition to this the study also ensures that the existing literature when practically applied gives the same results so the study tested the literature with its application as despite extensive growth of ERP systems ERP research area is still lacking in filling the gap between its literature and its application in practical (Al-Mashari, 2003) . The paper is sequenced as the literature review of ERP systems, research study, Case study through exploration of the hypothesized crucial factors, hypotheses testing through data analysis and at the end findings and conclusions.

**LITERATURE REVIEW**

ERP systems have been the most popular IT systems, which have been adopted by both private and public sector organizations (Davenport, 2000). Legacy systems are replaced primarily for cost reduction (Davenport T. H., 1998), (Mabert et al, 2003) by MNEs in the long run but ERP projects are handled in a different manner as other information technology systems are implemented (Davenport, 2000) because of the width of ERP systems implementation that covers not only managerial and operational constructs components but strategic and organizational constructed components (Markus et al, 2000), (Yu, 2005). The study focused on five major crucial factors, which ultimately have an impact on ERP system performance in an organization.

**Business Process Re-engineering:** ERP systems are developed based on the best industry practices (O'Leary, D. E.,2000) but not necessarily suits every organization's business processes. Same ERP product may result differently in two similar organizations in the same industry even. A process-IS/System mismatch is avoided through software customization or process reengineering. ERP unlike other information systems or software reposition its implementing company and transform business practices (Jarrar et al, 2000). Organizations either go for change in business processes to be aligned with the ERP systems or ERP packages customization is made (Jarrar et al, 2000) which may result in higher cost and longer time for its implementation ((Davenport, 1999). Organizations with ERP implementation in pipeline focus towards BPR by reformulating their exiting processes towards the standardized ones. Organizational readiness for the change in business processes ensures the successful accomplishment of BPR (Zhang et al, 2002), which ultimately lead it towards ERP system success. Different research studies agreed on BPR for successful ERP implementation but differ on its time. It suggested BPR before or through implementation (Bancroft et al,1998; Somers et al, 2004) where as Welti N. (1999), supported after the project implementation due to the point of functionality of the software not known in earlier phases.

**Architectural choices:** ERP systems carry high level of risks especially high cost as millions of dollars are spent on ERP purchase and its implementation but to an extent avoid spending even a small fraction of it on investigation of the best suited one from different available options (Al-Sehali, S., 2000). Study shows that most of the ERP systems are backed by advanced components of technology as GUIs, Relational Databases, high level forth generation languages CASE tools, and client server architecture (Anderegg et al, 2000). Looking into the deep of available options some ERPs may have certain limitations and portfolios like these may be industry specific, inflexible or rigid etc. Different ERP systems may have success stories in selected industries or in selected markets (Akkermans, H. and Helden, K., 2002).

Organizations should go for the ERP systems/software keeping their hardware/software infrastructure, its databases and its operations in consideration (Zhang et al, 2003). The basic need of investigation is software/hardware compatibility and software/application flexibility assurance in case customization is required. In addition to that once the product is decided, have to investigate further about the software versions and modules that best suit the organization. A wrong assessment in this regard may result in both high risk of time and cost overrun (Akkermans, H. and Helden, K., 2002; Janson et al, 1996). The whole ERP package, if rightly selected in the initial phases of its implementation lifecycle, provides the exact required track to the whole project (Somers et al, 2004). Despite having a track of right package selection, which should be intertwining organizational needs, is quite critical (Gattiker et al, 2002) but ensures relatively greater chances of project success (Somers et al, 2004; Al-Mashari, 2003) and process improvement.

**Effective usage of process database:** Previous research studies put intense emphasis on process database through in depth analysis of different ERP projects for their successful implementation. Especially at the times of planning and productivity analysis of ERP projects (Grady et al, 1999). The study when drilled down the existing literature derived and output that is the data used for legacy systems may cause hurdles in an ERP project success (Gattiker et al, 2002). Data with flaws creates critical problems for the ERP implementation and this problem crawls from one module to the others (Umble et al, 2002; Zhang et al, 2002). PTCL though transformed its data into the required ERP input format as literature reveals Reif, H. (2001).

**Education on new business processes** New business processes remained the foremost priority of the organizations where ERP implementation made. Organizations generally spend varying amount of user training. It may be 50% (Davenport, 2000) or up to 5% (Wheatley, 2000) of the total cost. Educating the internal stakeholders about new business processes, which ultimately are to be addressed through any of the ERP modules, are generally underestimated from budget side. This approached is which is technically wrong because the project goes towards failure if the knowledge workers are not educated about the new set of business processes in addition to the training of the software interfaces. Investment of education of the users is responsive (Sumner, 1999) and leads ERP project towards its success

**Technical selection of quality consultant:** Selection of a consultant in ERP projects plays a very critical role because of its services, which fills the gaps by ensuring the application of its timely expertise and out of the box thinking (Khan, 2002). Quality of the consultant is reflected through its profile and in depth knowledge base of the software (Welti, 1999; Al-Sehali, 2000). Selection of a quality consultant always remained critical for ERP projects especially the strategies and agreements placement. A quality consultant generally behaves as a key to the project and behaves as a fast driver of the project, which on the other way around put the project at stake and leads the project towards decline. Organizations with ERP implementation in pipeline or on papers to adopt it despite all the competencies and skills need a quality consultant. The studies show that in order to achieve success in ERP projects IT vendor/consultant is employed (Thong et al,1994; Janson et al,1996: Willcocks et al,2000). A tricky and complex exercise of selection of a quality consultant at times paves the way for the consultant to have project ownership, which ultimately is like aids for the ERP implemented company. In the past as study explores that most of the organizations with ERP implementation the quality of the consultants din not proved remarkable which ultimately resulted in project and organization collapse (Markus et al, 2000). Having a quality consultant doesn't mean to hand over project ownership to it but their skills and knowledge be utilized to the full because it costs very high to the business (Skok et al, 2002).

Study proposes a conceptual model, which indicates the five most critical factors. When the literature review has been explored the conceptual model with five crucial technical factor found dominatingly on ERP performance i.e. Fig 1.1. But when these factors have been applied on the real world scenario it is proved that these factors not only contribute in performance improvement but they all have a very strong correlation amongst themselves as well as shown in Fig 1.2. The study also intends to carry this research forward in future by extending the current model of ERP Performance enhancement, which should be covering more technical crucial factors.

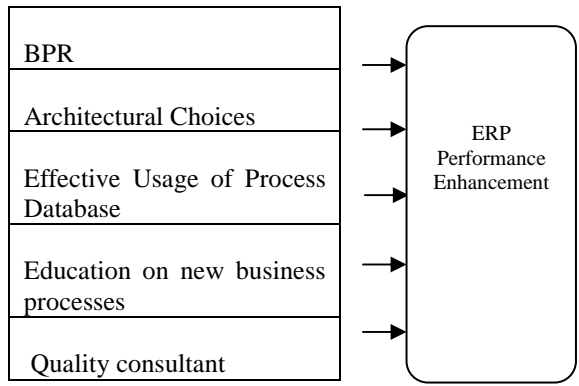

Figure 1.1 ERP Performance Enhancement Conceptual Model.

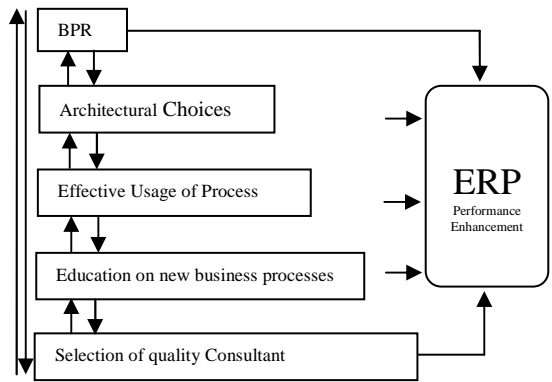

Figure 1.2. ERP Performance Enhancement Conceptual Model

**RESEARCH STUDY**

**Hypothesis 1 (H1):** BPR is better option for a public sector organization as compared to the customization in ERP software.

**Hypothesis 2 (H2):** Architectural choices lead to software process improvement for ERP projects.

**Hypothesis 3 (H3):** Process database management proved effective for successful ERP implementation.

**Hypothesis 4 (H4):** Higher the level of education on new business processes, higher the performance benefits of ERP systems would be.

**Hypothesis 5 (H5):** Proactive approach of selecting consultant for ERP implementation raises the performance of ERP systems.

**EXPLORATORY CASE STUDY**

In order to carry forward the research of the previous scholars and in addition to the literature review, a case study was conducted at PTCL, a largest telecom services provider in Pakistan. Initially some informal interviews were conducted from the technical people and the people from the management, who had been involved in SAP implementation in the company. On the basis of highlighted issues from these interviews a questionnaire was formed contains 23 questions, which ultimately brought the study towards five variables that are the most critical factors of performance improvement of SAP implementation project. PTCL has established a separate department of ERP with employee strength of around 58. These 58 people are working on ERP system and being trained along side of their routine work. In addition to these employees there are hundreds of employees who have been involved in ERP implementation from almost all the departments. This survey though was conducted by having random samples amongst all the above-mentioned employees because of the reluctance of most of the employees to respond. The study restricted itself on to five factors as mentioned in the literature review section because of the intensive feed back of the respondents from PTCL. So the study finds itself bound under the limitations of short time and the respondents' reluctance because of the threats of being politicized or the threats of jobs. However endeavors have been put together to explore these five technical factors to their full.

**5. HYPOTHESES TESTING, DATA ANALYSIS AND FINDINGS:**

After the completion of survey, data from the questionnaires was interpreted for each developed hypothesis. Cronbach Alpha derived from the data was 77.5%, which shows reliability of the questions. After ensuring the reliability of the questionnaire all the questions were tested. One sample t-test when applied resulted in rejection of all the null hypothesis developed in the study. Level of significance i.e. .05, ensures that the sample mean is greater then all the five variables, which definitely satisfy the acceptance of asked questions from the respondents. The provided tables provide the realistic analysis of what has been stated in both the models i.e. Fig 1.1 & Fig 1.2. Results are summarized for five hypothesis mentioned above in Table 1, Table 2, Table 3, Table 4, and Table 5. Empirical results shown in these tables are highly significant and reject the all five null hypothesis. The results are in support of alternate hypothesis.

On the basis of the data analysis, study confirms that the addressed crucial technical factors would be enhancing the performance of ERP projects along with rest of the researched managerial and other technical factors. In addition to the survey questions, the informal interviews helped the study to confirm that PTCL suffered with the said technical problems and paid a heavy cost for it. Though PTCL took care of all these crucial factors but not in an ideal manner and these were the core reasons where project performance deteriorated. Education on business processes and Quality vendor selection carry both managerial and technical perspectives, if any of these is missing the other would be useless.

The study proves in both the cases i.e. literature review and the case study method that these unique and most crucial factors ultimately become a reason of performance improvement and deterioration of ERP projects. The cited factors though need different managerial inputs but the technical inputs have been proved must for the overall projects' performance enhancement. Market competition amongst different ERP vendors is getting tough day by day because of the best industry practices embed in their product as SAP, Oracle Financials, BAAN, JD Edwards etc but the most fittest is the one which need minimal customization to best address the companies' business processes or can productively support the company in developing new business processes to get best out of their software. In the phase of planning different types of analysis guide the company to get the answers of 'What', 'Why', 'How' and 'When' for reliable decisions regarding ERP implementation. When necessary considerations left missing like the cited ones because of any good or bad reason, the project suffers much. Business Processes Change is considered as the most critical method in any company because of carrying high risk factors. So mostly companies try to customize the software to make them behave as the existing process require. In support of this argument configuration of the business requirements into the software is a highly complex exercise which most often results in time and cost overrun and may result in project failure or at least performance decrease.

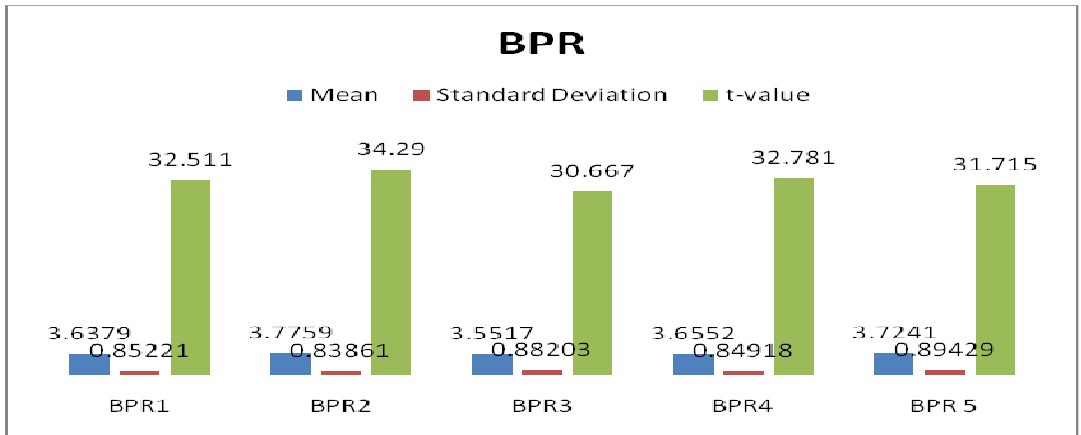
**Fig 2 Results of Hypothesis 1**

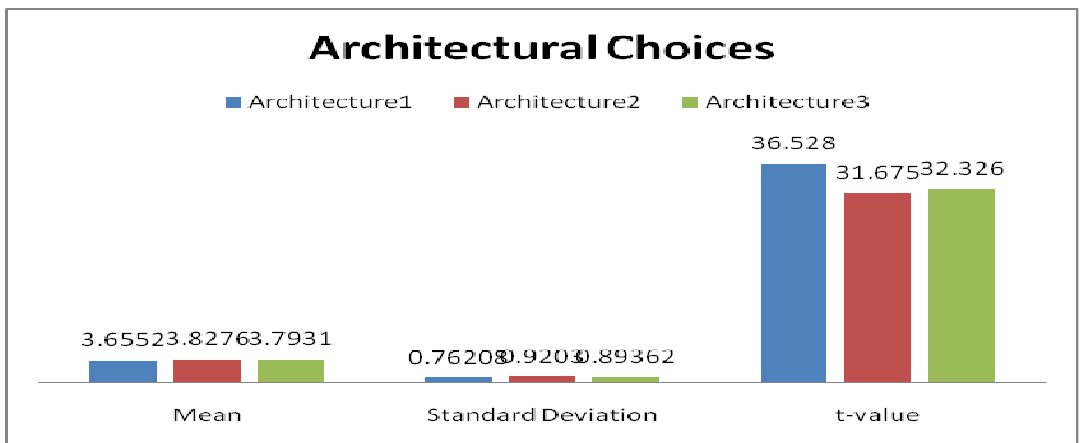
**Fig 3 Results of Hypothesis 2**

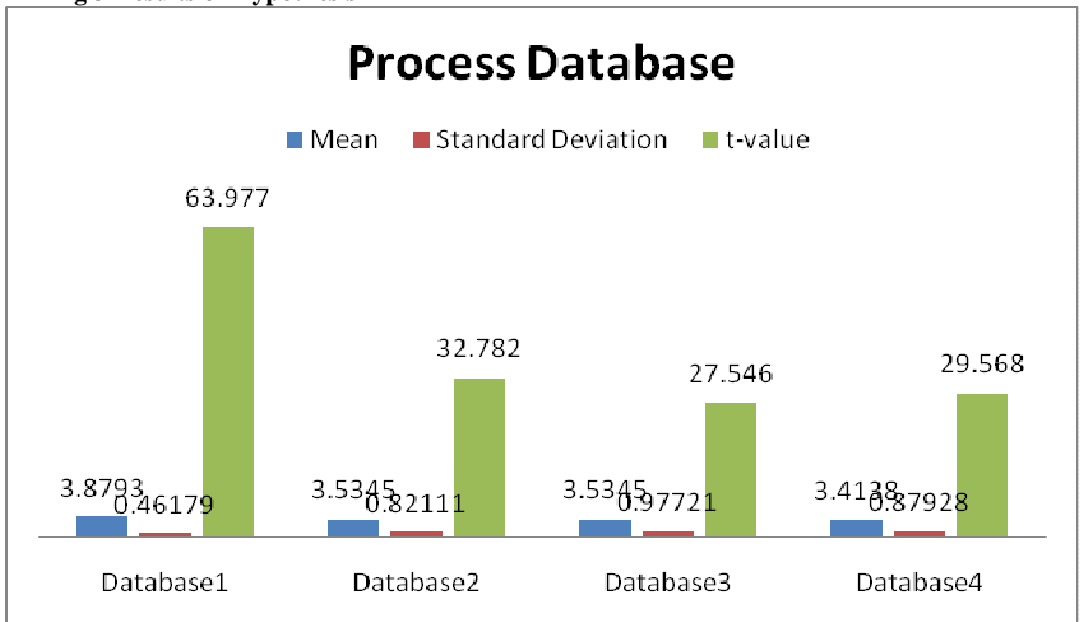
**Fig 4 Results of Hypothesis 3**

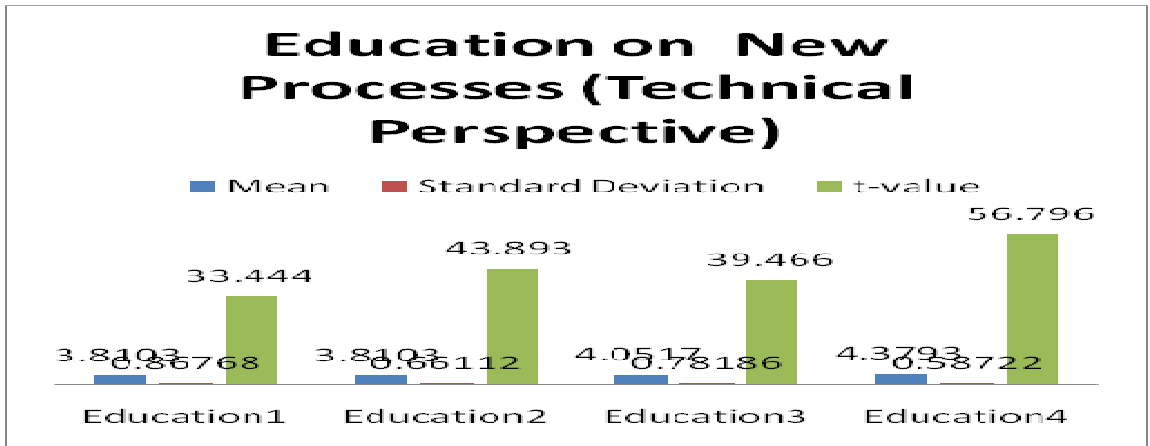
**Fig 5 Results of Hypothesis 4**

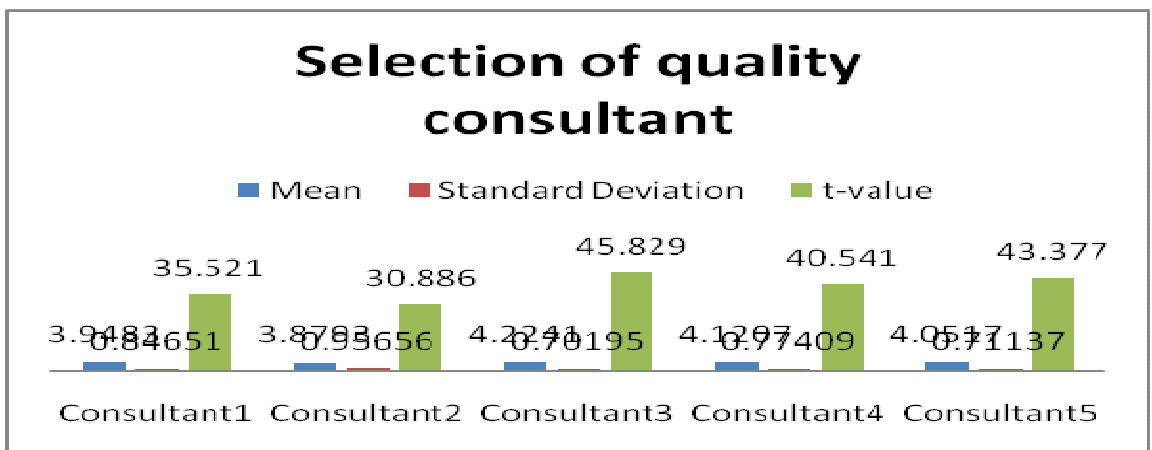
**Fig 6 Results of Hypothesis 5**

**CONCLUSIONS**

Out of the results drawn from the data analysis carry the study towards the goodness of BPR. According to the study SAP ERP software is developed, based on the industries' best practices with its deep breadth of offered processes. Out of the detailed literature review the same conclusion was drawn. Though the literature explores performance enhancement factors in detail especially the case studies but it seems that there is a gap between the actual happenings and the literature. Research in the past mainly focused on the management side or the critical success factors from management point of view but it lacks technical perspective, which should include the selected factors.

**Table 1: Results of Hypothesis 1**

| Survey Questions | Mean | Standard Deviation | t-value | P (t-value) for one-tail test | Results |
|---|---|---|---|---|---|
| BPR1 | 3.6379 | .85221 | 32.511 | .000 | Rejects Null Hyp. Support H1 |
| BPR2 | 3.7759 | .83861 | 34.290 | .000 | Rejects Null Hyp. Support H1 |
| BPR3 | 3.5517 | .88203 | 30.667 | .000 | Rejects Null Hyp. Support H1 |
| BPR4 | 3.6552 | .84918 | 32.781 | .000 | Rejects Null Hyp. Support H1 |
| BPR 5 | 3.7241 | .89429 | 31.715 | .000 | Rejects Null Hyp. Support H1 |

**Table 2: Results of Hypothesis 2**

| Survey Questions | Mean | Standard Deviation | t-value | P (t-value) for one-tail test | Results |
|---|---|---|---|---|---|
| Architecture1 | 3.6552 | .76208 | 36.528 | .000 | Rejects Null Hyp. Support H1 |
| Architecture2 | 3.8276 | .92030 | 31.675 | .000 | Rejects Null Hyp. Support H1 |
| Architecture3 | 3.7931 | .89362 | 32.326 | .000 | Rejects Null Hyp. Support H1 |

**Table 3: Results of Hypothesis 3**

| Survey Questions | Mean | Standard Deviation | t-value | P (t-value) for one-tail test | Results |
|---|---|---|---|---|---|
| Database1 | 3.8793 | .46179 | 63.977 | .000 | Rejects Null Hyp. Support H1 |
| Database2 | 3.5345 | .82111 | 32.782 | .000 | Rejects Null Hyp. Support H1 |
| Database3 | 3.5345 | .97721 | 27.546 | .000 | Rejects Null Hyp. Support H1 |
| Database4 | 3.4138 | .87928 | 29.568 | .000 | Rejects Null Hyp. Support H1 |

**Table 4: Results of Hypothesis 4**

| Survey Questions | Mean | Standard Deviation | t-value | P (t-value) for one-tail test | Results |
|---|---|---|---|---|---|
| Education1 | 3.8103 | .86768 | 33.444 | .000 | Rejects Null Hyp. Support H1 |
| Education2 | 3.8103 | .66112 | 43.893 | .000 | Rejects Null Hyp. Support H1 |
| Education3 | 4.0517 | .78186 | 39.466 | .000 | Rejects Null Hyp. Support H1 |
| Education4 | 4.3793 | .58722 | 56.796 | .000 | Rejects Null Hyp. Support H1 |

**Table 5: Results of Hypothesis 5**

| Survey Questions | Mean | Standard Deviation | t-value | P (t-value) for one-tail test | Results |
|---|---|---|---|---|---|
| Consultant1 | 3.9483 | .84651 | 35.521 | .000 | Rejects Null Hyp. Support H1 |
| Consultant2 | 3.8793 | .95656 | 30.886 | .000 | Rejects Null Hyp. Support H1 |
| Consultant3 | 4.2241 | .70195 | 45.829 | .000 | Rejects Null Hyp. Support H1 |
| Consultant4 | 4.1207 | .77409 | 40.541 | .000 | Rejects Null Hyp. Support H1 |
| Consultant5 | 4.0517 | .71137 | 43.377 | .000 | Rejects Null Hyp. Support H1 |

The researched factors in the paper proved to be the most critical ones especially when applied to the public sector organization in Pakistan. Public sector set of processes differs generally from the private sector implementation. These are somewhat rigid and require more in these critical factors. The focused technical factors provide ERP projects avoidance from the failure and at later stages, if addressed properly drive the project towards overall performance enhancement. These critical technical factors have been proved from the existing literature and the real application in telecom sector of Pakistan, which has been authenticated through the data analysis. Initially an in depth literature review was made to best analyze the addressed technical critical factors. Then the same technical factors were applied in PTCL where ERP implementation is in progress through another case study methodology. Architectural choices being one of the most technical critical factors if addressed properly saves the life of the ERP project along with the process database, new process training and consultant selection and management. All these technical factors are interdependent on each other as well. The less focus on one may cause the negative effect on the other and ultimately project suffers. Study proves that though managerial focus should be on priority but absence of these factors never let the management go smooth in ERP projects. Paper following two methods i.e. literature review and case study method reveals that these five technical factors can enhance the performance of ERP projects but at the same time destroy the whole project if not properly addressed. These factors are interdependent on each other, which drive their application in a parallel manner. Still many other technical aspects are needed to be explored which will be the part of future research in continuation of this study.